\documentclass[preprint,prb]{revtex4}

\usepackage[ps2pdf,bookmarks=false,urlcolor=red,citecolor=blue,linkcolor=blue,
              pagecolor=red,hyperindex,colorlinks,hyperfigures%pagebackref=true
        ]{hyperref}
\usepackage{amsmath, amssymb, amsthm}
\usepackage{graphicx}
\begin{document}
%\begin{CJK*}{GBK}{song}

\title{Tunable Localization and Oscillation of Coupled Plasmon Waves in Graded Plasmonic Chains}

\author{M. J. Zheng,$^{1}$ J. J. Xiao,$^{2,3}$ and K. W. Yu$^{1,}$\footnote{Electronic mail: kwyu@phy.cuhk.edu.hk} }
%\author{M. J. Zheng (\textbf{郑明杰}),$^{1}$ J. J. Xiao (\textbf{肖君军}),$^{2,3}$ and K. W. Yu (\textbf{余建华})$^{1,}$\footnote{Electronic mail: kwyu@phy.cuhk.edu.hk} }

\affiliation{$^1$Department of Physics and Institute of Theoretical Physics, The Chinese University of Hong Kong, Shatin, N. T., Hong Kong, China\\
$^2$Department of Physics, The Hong Kong University of Science and Technology, Clear Water Bay, Hong Kong, China\\
$^3$Department of Electronic and Information Engineering, Shenzhen Graduate School, Harbin Institute of Technology, Shenzhen 518055, China}

\begin{abstract} The localization (confinement) of coupled plasmon modes, named as gradons, has been studied in metal nanoparticle chains immersed in a graded
dielectric host. We exploited the time evolution of various initial wavepackets formed by the linear combination of the coupled modes. We found
an important interplay between the localization of plasmonic gradons and the oscillation in such graded plasmonic chains. Unlike in optical
superlattices, gradient cannot always lead to Bloch oscillations, which can only occur for wavepackets consisting of particular types of
gradons. Moreover, the wavepackets will undergo different forms of oscillations. The correspondence can be applied to design a variety of
optical devices by steering among various oscillations.
\end{abstract}

\date{\today}

\maketitle

\newpage

\section{Introduction}

Clusters of metal particles at the nanometer scale have attracted much interest because of their plausible applications in
nano-optics.~\cite{Phys.Rep.431.87} The main objective is to overcome the diffraction limit, achieving subwavelength confinement of waves by
surface plasmons.~\cite{Science311.189} In the last decade, enormous developments have been witnessed in plasmonic waveguides to build
all-optical circuits.~\cite{Science311.189, JApplPhys.011101} Regarding controlled localization of optical signals and guided electromagnetic
energy at subwavelength scale, there have been various progresses.~\cite{PhysRevLett.91.227402, PhysRevLett.93.137404, PhysRevB.70.125429}
Furthermore, Fabry-Perot cavity was proposed to confine surface plasmons.~\cite{Appl.Phys.Lett.86.161101} Light will be localized when the
frequency matches the surface plasmon eigenfrequency of the system, while for a wavepacket consists of a variety of frequencies, oscillations
will occur.

The most intriguing oscillation is the Bloch oscillation (BO), which is the oscillatory motion of a particle in a periodic potential when a
constant force is acting on it. It has been generally understood that Bragg reflections together with a linear gradient of the potential can
cause Bloch oscillation of a wave of distinct nature (electronic, optical, acoustic, or matter wave).~\cite{PhysRevLett98.134301} The gradient
in potential can be introduced by an external field or a spatial perturbation of any nature (electric, magnetic, acceleration, or gravitation
field).~\cite{PhysRevLett98.134301} For example, in optical structures, the optical equivalent of an external field in electronics can be a
gradient in the refractive index \cite{PhysRevLett91.263902} or a geometrical variation in waveguides.~\cite{PhysRevB.76.195119,
OptLett.32.2647} Many investigations have been devoted to photonic BO in helical (deformed) waveguides,~\cite{PhysRevB.76.195119} curved
waveguides,~\cite{PhysRevLett.83.963} waveguide lattices,~\cite{JOptSocAmB.24.2632} chirped metamaterials,~\cite{OptExpress.16.3299} and other
photonic heterostructures.~\cite{RevModPhys.78.455} Recently, plasmonic Bloch oscillations (PBO) have been observed in metal heterowaveguide
superlattices,~\cite{ApplPhysLett.91.143121.2007} metal waveguide array structures,~\cite{ApplPhysLett.91.243113.2007} and chirped
metal-dielectric structures.~\cite{ApplPhysLett.94.161105.2009} However, the mechanisms of BO, especially the relation between the occurrence of
BO and the condition of gradient, have not been clearly explained. In this work, we aim to explore the mechanisms of dynamic oscillations
(including BO) in graded plasmonic chains (GPCs) based on our understanding of localized plamonic modes (gradons) in previous
work.~\cite{ApplPhysLett88.241111, ApplPhysLett89.221503}

In GPCs, the gradient can be realized by gradually changing the dielectric constant of the host,~\cite{ApplPhysLett88.241111,
ApplPhysLett89.221503} the distance between the metal particles, and/or the particle size along a particular spatial
dimension.~\cite{NanoLett8.2369, PhysicaB394.208} There exist peculiar confined eigenmodes (or quasi-normal modes) which are called gradons in
the graded lattices.~\cite{ApplPhysLett88.241111, ApplPhysLett89.221503} Gradons  with different frequencies are localized at different position
and have different spatial extent in the system. Then one would ask the following questions: Can BO occurs in a GPC? What is the mechanism of BO
in graded optical or plasmonic lattices? Does a gradient guarantee the occurrence of BO? Motivated by these questions, we aim to explore the
relation between gradon localization and the occurrence of BO in GPCs. For this purpose, we shall use a phase diagram to distinguish between
various plasmonic gradon modes in GPCs. Then through a dynamic simulation, we show the evolutions of different ``wavepackets'' in the
position-time domain. We find that there exists a clear and interesting correspondence between the gradon localizations and the occurrence of
BO, breathing-wave-like oscillations, or other kinds of oscillations. We explicitly show that only initial ``wavepackets'' formed by a
combination of certain type of gradons  can undergo BO or breathing-wave-like oscillation. The existence of such kind of gradons requires rather
strong gradient strength and they are referred as blue-red gradons which are well confined in the ``internal'' region of the system. Conversely,
BO and breathing-wave-like oscillation cannot occur in GPCs bearing no such blue-red gradons, irrespective of whatever initial ``wavepacket''.
Despite of observing BO in GPCs, we further explore its mechanism and find that gradient can not always lead to BO in GPCs. Our results provide
useful guidelines in designing optical devices of novel functionalities.

\section{Model and formula}

We consider a chain of $N$ spherical silver nanoparticles that is immersed in a dielectric
host.~\cite{ApplPhysLett88.241111,ApplPhysLett89.221503} The dielectric function of the metal nanoparticles has the generalized Drude form
\cite{PhysRevB71.045404} $\epsilon _m (\omega )=\xi -\eta \omega _p^2 /\omega (\omega +i\Gamma )$, while the host has a graded dielectric
constant varying along the chain axis, i.e., $\hat {x}\mbox{,}$ as $\epsilon_h (x_n )=\epsilon_l +cx_n /l \quad (n=1,2,...,N)$, where $x_n =nd_0
$ is the position of the $n$th nanoparticle, $d_0$ the interparticle spacing, $l=Nd_0 $ the total length of the chain, and $c$ is the
coefficient of dielectric gradient. All the particles are identical in size with radii $a$. The $n$th and the $m$th particles are at a
center-to-center distance $d_{nm} =d_0 \vert n-m\vert $. The coupled equation for point dipoles reads~\cite{ApplPhysLett89.221503}
\begin{equation}\label{eq1}
{\rm {\bf G}}(\omega )\left| \phi \right\rangle =0,
\end{equation}
where $\left| \phi \right\rangle $ is the column vector of the dipole moments oscillating with frequency $\omega$, $G_{nn} (\omega )=1/\beta _n
(\omega )$ and $G_{nm} (\omega )=-\mathop {\tilde {T}}\nolimits_{nm} $ $(m\ne n)$. Here $\beta _n (\omega )=a^3\epsilon_h (x_n )[\epsilon_m
(\omega )-\epsilon_h (x_n )]/[\epsilon_m (\omega )+2\epsilon_h (x_n )]$ represents the polarizability of the $n$th particle, and $\mathop
{\tilde {T}}\nolimits_{nm} $ is the dipole electromagnetic interaction between the $n$th particle and the $m$th particles,
\begin{equation}\label{eq2}
\mathop {\tilde {T}}\nolimits_{nm} =\frac{1}{\epsilon_h (x_n )}(\nabla _n \nabla _n +k^2{\rm {\bf I}})\frac{e^{ik\vert x_n -x_m \vert }}{\vert
x_n -x_m \vert }\,,
\end{equation}
where $k$ is the wave vector in the host. We can take the point-dipole approximation in the calculation, because the condition $a/d\leq 1/3$ is
always satisfied. The validity of this approximation has been proved theoretically and experimentally.~\cite{NanoLett8.2369} To calculate the
eigenmodes of the system, we assume the external field is zero and take the wave vector $k = 0$. In the quasistatic limit, $\mathop {\tilde
{T}}\nolimits_{nm} \approx \kappa /\epsilon_h (x_n)d_{nm}^3$, where $\kappa =2(-1)$ for the longitudinal (transverse) case.

Actually, the generalized Drude model can be extended from the standard Drude model by replacing $\omega _p^2 $ with $\eta \omega _p^2 /\xi $
and $\epsilon_h $ with $\epsilon_h /\xi $, respectively. However, it is hard to determine the eigenvalues and eigenvectors by directly
diagonalizing ${\rm {\bf G}}(\omega )$, because Eq.~(\ref{eq1}) is a nonlinear complex transcendental equation with respect to the frequency
$\omega $. This prevents us from constructing a phase diagram which is crucial to understand the connection between various gradon modes and the
dynamic oscillations. In order to circumvent the difficulties, we linearize Eq.~(\ref{eq1}) with respect to the frequency by using the
quasistatic approximation of Eq.~(\ref{eq2}). In this way, we can use standard diagonalization to obtain both the eigenvalues and eigenvectors.
For noble metal, e.g., silver, $\Gamma \approx 0.005 \omega_p$ according to the tabulated data in the reference.~\cite{Handbook1985} Here we can
neglect the loss, i.e., $\Gamma =0$ in the metal nanoparticles. Thus Eq.~(\ref{eq1}) can be linearized with respect to $\omega ^2$
\begin{equation}\label{eq3}
({\rm {\bf F}}-\omega ^2{\rm {\bf I}})\left| \phi \right\rangle =0\,,
\end{equation}
where ${\rm {\bf I}}$ is an identity matrix and $F_{nm} =\omega _n^2 -G_{nm} (\omega _n^2 )/G_{nn}^\prime (\omega _n^2 )$. Here the prime
indicates derivative with respect to $\omega ^2$ and $\omega _n =\omega _p \sqrt {\eta /(\xi +2\epsilon_h (x_n ))} $ is the dipole Mie resonant
frequency of the $n$th particle at that particular position where the host has dielectric constant $\epsilon_h (x_n )$. Equation~(\ref{eq3})
represents a very good approximation of Eq.~(\ref{eq1}) when $\omega \approx \omega _n$ and it can be mapped onto an equivalent elastic chain
with mass $M_n =[\xi +2\epsilon_h (x_n )]^2/3\lambda \eta \omega _p^2 \epsilon_h (x_n )$, and additional on-site potential $U_n =M_n \omega _n^2
-2K_0 $, where $K_0 =(a/d_0 )^3$ is a force constant between adjacent masses.~\cite{ApplPhysLett88.241111, ApplPhysLett89.221503} Two
characteristic frequencies $\omega _{c1} (n)=\sqrt {U_n /M_n }$ and $\omega _{c2} (n)=\sqrt {(U_n +4K_0 )/M_n}$ were defined to construct the
phase diagram.~\cite{ApplPhysLett89.221503} The parameters used in this work are listed below: $N=100$, $c=1.0$, $\epsilon_l =3.0$, $\xi =5.45$,
and $\eta =0.73$. Thus the host dielectric constant varies from $\epsilon_h ($1$)=3.0$ (left end) to $\epsilon_h ($N$)=4.0$ (right end).

\section{Plasmonic gradon modes in graded plasmonic chain}

Here we briefly summarize the various localized plasmonic gradon modes and their transitions,~\cite{ApplPhysLett89.221503} which are very
related to the dynamic oscillations of light wavepackets in the graded plasmonic chains. The effective approach of identifying various gradon
modes is the phase diagram,~\cite{JPCM19.026224} as shown in Fig.~\ref{fig:gpc:PD}, which contains four regions representing four different
kinds of gradon modes. The different phase regions are interfaced by four curves denoting, respectively, $\omega_{c1}(1)$ (dashed line),
$\omega_{c1}(N)$ (solid line), $\omega_{c2}(1)$ (dotted-dashed line), and $\omega_{c2} (N)$ (dotted line) as functions of distance $d_0$. In
each region, there exists a kind of plasmonic gradon modes, which is localized at different position of the system. The mode patterns of some
plasmonic gradon modes, i.e. square moduli of induce dipoles versus position, are shown in the insets. There exists a critical value $d_0=d_c$,
where $\omega_{c1}(1) = \omega_{c2}(N)$. For interparticle spacing $d_0 < d_c$, the unbound (extended) modes (U) cover a frequency range
$\omega_{c1}(1)<\omega <\omega_{c2}(N)$ (vertical line shaded region). In the lower frequency range $\omega _{c1} (N)<\omega <\omega _{c1}(1)$
(when $d_0 < d_c$) and $\omega _{c1} (N)<\omega<\omega_{c2}(N)$ (when $d_0 > d_c$), the modes are localized at the right hand side of the chain,
and are called red plasmonic gradons (RG) (crossed shaded region). In the higher frequency range $\omega_{c2}(N)<\omega<\omega _{c2}(1)$ (when
$d_0 < d_c$) and $\omega_{c1}(1)<\omega<\omega_{c2}(1)$ (when $d_0 > d_c$), the modes are localized at the left hand side of the chain, and are
called blue plasmonic gradons (BG) (tilted dotted line shaded region). For interparticle spacing $d_0 > d_c $, the modes become localized in the
middle parts of the chain and covers a frequency band $\omega _{c2} (N)<\omega <\omega _{c1} (1)$ (wavy shaded region). We call these modes
blue-red plasmonic gradons (BRG) which are of great significance for the oscillations described below. Since the evolution of incoming wave
depends on the initial conditions, we need to do the following wavepacket dynamic analysis.

\section{Wavepacket dynamic analysis of various oscillations of coupled plasmon waves}

Once the complete set of gradon eigenmodes $\left| \phi \right\rangle $ has been obtained, we can construct an initial state to perform time
domain simulations on the ``wavepacket'' dynamics. Let us take an initial wavefunction,
\begin{equation}
\label{eq4} \psi (x,0)=\frac{1}{(2\pi \sigma _x^2 )^{1/4}}\exp {\left[{-\frac{(x-x_0 )^2}{4\sigma _x^2 }} \right ]} e^{-ik_0 x},
\end{equation}
where $k_0$ is the vacuum wavenumber determining the central frequency and the direction of propagation of the wavepacket. The intensity profile
$\vert \psi (x,0)\vert ^2$ has a Gaussian distribution centered at $x_0 $ with spatial width $\sigma _x $. We then expand the initial
wavefunction in terms of $\left| \phi \right\rangle $,
\begin{equation}\label{eq5}
| \psi (0)  \rangle =\sum\limits_n A_n | \phi_n \rangle \,,
\end{equation}
where $A_n = \langle \phi_n|\psi(0)\rangle$ is the constituent component of the initial wavepacket. The intensities $\vert A_n \vert ^2$ of the
various initial wavepackets have been shown as insets in Fig.~\ref{fig:gpc:CONTOUR}. The peak of $\vert A_n \vert ^2$ corresponds to the
contribution from a dominant component of frequency $\omega _n $, which is in agreement with the central frequency obtained from the dispersion
relation $\omega (k)$ at $k_0 $. Thus the subsequent wavefunction at time $t$ is,
\begin{equation}\label{eq6}
| {\psi (t)} \rangle =\sum\limits_n A_n | {\phi_n} \rangle e^{i\omega _n t}.
\end{equation}
When the size $N$ becomes large, one can replace the sum by an integral over the spectral components in Eqs.~(\ref{eq5}) and (\ref{eq6}). The
evolution of the wavepacket intensity $\vert \psi (x,t)\vert ^2$ can be used to illustrate the various oscillations, including BO,
breathing-wave-like (BW) oscillations and other kinds of motions. We will show that combination of different localized modes (gradons) lead to
different dynamic evolutions of the wavepacket.

Here a few cases will be addressed to illustrate the correspondence between the various gradons  and the oscillation of wavepackets. The
occurrence of various oscillations depends on the localization extents of modes forming the initial wavepackets. Since the phase diagram as
sketched in Fig.~\ref{fig:gpc:PD} explicitly reflects the spatial extension of the various gradon modes that the graded plasmonic chain can
sustain, it can help us to have an instant judgment that whether an initial wavepacket contains a component that can reach either the left or
the right end of the chain, where reflection occurs. Such kind of wavepackets shall not undergo Bloch oscillation or breathing-wave-like
oscillation. A necessary condition for occurrence of BO or BW oscillation is that the initial wavepacket only consists of blue-red gradons,
because BRG are localized in the middle part of the chain. We have marked the constituent components (intensity versus frequency) of the
different initial wavepackets at certain $d_0$ on the phase diagram (see Fig.~\ref{fig:gpc:PD}). The corresponding dominant gradon mode of the
initial wavepackets is shown in the insets, i.e. plots of $|\psi(x)|^2$ versus $x$. As expectations, in the BRG region, BO can occur. Depending
on the initial wavepacket, breathing-wave-like (BW) oscillations can also occur. The differences are that the initial wavepacket that undergoes
BO has a larger spatial width (narrow bandwidth) while the initial wavepacket that undergoes BW oscillation has a relatively smaller spatial
width (broad bandwidth). In the RG region, the initial wavepackets consisting of red gradons must be reflected at the right (red) end of the
chain, such kind of dynamic evolution is denoted by RR (right end reflection). While in the BG region, left (blue) end reflection denoted by LR
occurs. Finally, for the U region, an initially extended wavepacket can be reflected by both ends of the chain, which is denoted by LRR (left
and right ends reflection).

To show more clearly the above analysis regarding gradon localization and the occurrence of BO, we compose the contour plots of the intensity
profile $\vert \psi (x,t)\vert ^2$ in the position-time (i.e., $x$-$t$) domain for various initial wavepackets. The white color and red color
indicate the very strong and relatively strong intensity, respectively, and the black color means the intensity is weak or zero.
Figure~\ref{fig:gpc:CONTOUR}(a) shows the evolution of $\vert \psi (x,t)\vert ^2$ for $d_0 =5.5a$, $\sigma _x =5$, and $k_0 =0.8\pi /d_0 $. The
wavepacket exhibits an oscillatory motion: the mean position shows a periodic time-dependence while the width is nearly constant. This is a
typical plasmonic BO process. More interestingly, in Fig.~\ref{fig:gpc:CONTOUR}(b), we show the case of $d_0 =5.0a$, $\sigma _x =0.2$, and $k_0
=0.8\pi /d_0 $. Now the wavepacket's width shows a periodic time-dependence but the mean position is nearly fixed at the initial place. This is
a breathing-wave-like (BW) oscillation. Both initial wavepackets for BO and BW oscillation are only formed by blue-red gradons which are normal
modes localized in the middle part of the graded chain. The typical modes at central frequencies $\omega =0.243\omega _p$ (BO) and $\omega
=0.245\omega _p$ (BW) are shown in the insets of Fig.~\ref{fig:gpc:PD}. However, the spatial widths of the two initial wavepackets are
significantly different. For BO, the spatial width is larger than that of breathing-wave-like oscillation: the spatial width is $\sigma _x =5$
for BO and $\sigma _x =0.2$ for BW.

We can estimate the period of Bloch oscillation. In semi-classical theory,~\cite{PhysRevLett98.134301} the equation of motion of the Bloch
wavenumber $k$ reads $\dot {k}=-\partial \omega /\partial x=f$, where $f$ is almost constant for BO. The time taken for $k$ to change by $2\pi
/d_0 $ is defined as the period of BO, denoted as $T_B$:
\begin{equation}\label{eq7}
T_B =\frac{2\pi }{f}\approx \frac{2\pi N(2\epsilon_l +c+\xi)^{3/2}}{c\sqrt \eta \omega _p }.
\end{equation}
Here the simplification is obtained by the fact that the average of $\cos k$ vanishes in a period. The plasmon resonant
frequency~\cite{PhysRevB71.045404} is $\omega _p =1.72\times 10^{16}$ rad/s. Thus the period of BO in Fig.~\ref{fig:gpc:CONTOUR}(a) is about
$11.8$ ps. This is in agreement with the previous work for optical BO.~\cite{PhysRevLett98.134301} For non-Bloch oscillations, the time between
two neighboring reflections is defined as the period $T_R$.

Despite of the oscillation dynamics, we have also investigated the non-Bloch oscillations, e.g. wave reflection from the left end, the right
end, or from both ends. Figure~\ref{fig:gpc:CONTOUR}(c) and (d) show the corresponding situations for reflection from the left and right end,
respectively. In Fig.~\ref{fig:gpc:CONTOUR}(c) ($d_0 =3.5a,\sigma _x =5,k_0 =0.9\pi /d_0 )$, the initial wavepacket is constructed by linear
combination of blue gradons only, which are localized modes residing at the left hand side of the chain (as shown in the inset named as LR in
Fig.~\ref{fig:gpc:PD}), this wavepacket can be reflected by the left end of the chain and can not reach the right end. In
Fig.~\ref{fig:gpc:CONTOUR}(d) ($d_0 =4.0a,\sigma _x =5,k_0 =0)$, the initial wavepacket is only formed by red gradons which are modes confined
at the right hand side of the chain  (as shown in the inset named as RR in Fig.~\ref{fig:gpc:PD}), the wavepacket is reflected at the right end
of the chain and can not reach the left end. If the components of initial wavepacket all fall into the extended modes region (see
Fig.~\ref{fig:gpc:PD}), the wavepacket can reach both ends and be reflected in multiple fashion before eventually spreading across the whole
chain (the contour plot is not shown here).

The different features of various oscillations can also be demonstrated in the contour plots of $\vert \psi (x,t)\vert ^2$ in reciprocal
position-time (i.e., $k$-$t$) domain. For BO or BW, $k$ varies periodically in the range $[-\pi,\pi]$, while for other kinds of oscillations,
$k$ varies only in part of the range $[-\pi,\pi]$. There are two reasons why we only show the contour plots of $\vert \psi (x,t)\vert ^2$ in
real spatial position-time (i.e., $x$-$t$) domain. One is the contour plots in real space can show the obvious features of BO, BW, or other
kinds of oscillations, but those in $k$ space only show the periodic variation of $k$ with time, which is sensitive to the initial value of $k$.
The second reason is that the difference between BO and BW is more obvious in real space than that in $k$ space.

\section{Evolution of wavepackets in damping case}

To be more realistic, we shall take into account the loss in the metal nanoparticles. For the damping case ($\Gamma \ne 0)$, the equation of
motion is modified by simply replacing $\omega ^2$ with $\omega (\omega +i\Gamma )$ that is used in Eq.~(\ref{eq3}). Thus the eigenvalues become
complex-valued $\tilde {\omega }=\omega _{0n} +i\gamma _n $, where $\gamma _n =\Gamma /2$ and $\omega _{0n} =\omega _n (1-\Gamma ^2/8\omega _n^2
)$. Thus the wavepacket is damped by an overall factor $e^{-\Gamma t/2}$. Figure~\ref{fig:gpc:EVOL} ($d_0 =5.5a$, $\sigma _x =5$, $k_0 =0.8\pi
/d_0$) shows the evolution of wavepackets for undamped ($\Gamma =0$, solid lines) BO in an ideal chain and damped ($\Gamma =0.08\omega _p$,
dashed lines) BO in silver nanoparticle chains. At $t=t_0$ (black lines) and $t=t_0 +0.5T_B$ (blue lines), the wavepacket reaches at the
right-most and left-most of the BO region. The damped wavepackets almost overlap the undamp ones. We can observe that the wavepacket at the
left-most is higher and narrower than that at the right-most. This is due to the width of wavepackets also oscillates with time, which is
similar to the photonic BO in electrically modulated photonic crystals.~\cite{OptLett.33.2200} As time goes on, at $t=t_0 +2.0T_B $ (red lines)
and $t=t_0 +4.0T_B$ (green lines), the intensity of damped BO becomes smaller than that of the corresponding undamp one, with the damping rate
$\Gamma /2$. At $t=t_0 +2.5T_B$ (orange lines) and $t=t_0 +4.5T_B$ (light green lines), the wavepacket of damped BO is higher and narrower than
that of corresponding undamp one. The amplitude of oscillation (i.e., distance between the left-most peak and right-most peak) is smaller in
damped case than that in undamp case.

\section{Discussion and further work}

In this work, we concentrate on studying the correspondence of gradon confinements and the Bloch oscillation as well as non-Bloch oscillations.
This is realized by applying a linear gradient potential in the periodic plasmonic chain. The linear gradient potential is brought by the graded
host permittivity. If we replace the linear gradient potential field by using a harmonic trap, we will realize the dipole oscillation. For
dipole oscillation, the momentum oscillates periodically in time at the bottom of the trap, the average of velocity is periodic in time, and the
phase difference between neighboring particles remain locked. However, dipole oscillation is unstable in optical waveguides with transverse
confinement. Long living dipole oscillations can be realized in photonic crystals with Kerr dielectrics by properly applying graded pump
fields.~\cite{OptLett.34.1777} Transition between Bloch oscillations and dipole oscillations will be studied in further work. Besides, based on
our understanding of transitions among different gradon modes and the correspondence between gradon confinements and various oscillations, we
can realize the steering among various oscillations by tuning the frequency or the graded parameters.

\section{Conclusion}

In conclusion, we elaborated the localized coupled plasmon modes and studied the dynamics of plasmonic wave in GPCs. We found that there exist
obvious correspondences between gradon localization and occurrence of various oscillations. The condition for occurrence of Bloch oscillation or
breathing-wave-like oscillation is that the initial wavepacket is formed only by blue-red plasmonic gradons. If the initial wavepacket consists
of other kinds of plasmonic gradon modes, e. g. blue (red) gradons, the wavepacket will be reflected by the left (right) end of the chain. The
predicted correspondence between gradon localization of initial wavepacket components and the type of subsequent dynamic oscillations opens the
possibility to design various optical devices. We employed the wavepacket dynamics here and exact numerical results were obtained. When the
chain size becomes large, some alternative methods, e.g., semi-classical theory and finite-difference time-domain methods can be used. The
evolution of damped Bloch oscillation in a realistic silver chain shows similar results as those of no damping case, except that the intensity
gradually decreases. The quasi-static dipole interaction has offered a good approximation, which gives similar results as that of fully retarded
interaction.~\cite{NanoLett8.2369} The marked phase diagram provides an effective tool for the experimental realization of Bloch oscillations
and for structure design of graded plasmonic arrays. A clear understanding of the mechanism of Bloch oscillations and other kinds of motions
enables us to make full use of the graded plasmonic arrays, as promising candidates for subwavelength optical circuit and optical storage
devices.

\hfill
\section*{ACKNOWLEDGMENT}

\noindent This work was supported by RGC General Research Fund of the Hong Kong SAR Government.

\newpage

\section*{Figure Captions}

\begin{figure*}[h t b p]\caption{(Color online) Phase diagram for the graded plasmonic chain ($N=100$, $c=1.0$, $\epsilon_l =3.0$, $\xi =5.45$, and $\eta =0.73$). There are four regions of different gradon modes in the phase
diagram as follows, BG: blue gradons region, RG: red gradons region, BRG: blue-red gradons region, and U: unbounded modes region. The
corresponding types dynamic motion types of wavepackets consisting of different gradon modes are named as follows, LR: left end reflection, RR:
right end reflection, BW: breathing-wave-like oscillations, and BO: Bloch oscillation. The typical mode patterns of these wavepackets are shown
in the insets. The extended mode in U region extends the whole system and is not shown here. Bloch oscillation and breathing wave-like
oscillation correspond to the same type of mode in the blue-red gradon phase, but occur with different initial wavepackets.} \label{fig:gpc:PD}
\end{figure*}

\begin{figure*}[h t b p]\caption{(Color online) Dynamics of various wavepackets shown by the contour plots of
$\vert \psi (x,t)\vert ^2$ on the $x$-$t$ domain. (a) Bloch oscillation (BO) ($d_0 =5.5a$, $\sigma _x =5$, $k_0 =0.8\pi /d_0 $, and $T_B =11.8$
ps), (b) breathing-wave-like oscillation ($d_0 =5.0a$, $\sigma _x =0.2$, $k_0 =0.8\pi /d_0 $, and $T_B =11.8$ ps), and (c) reflection from the
left (blue) end ($d_0 =3.5a,\sigma _x =5,k_0 =0.9\pi /d_0 $, and $T_R =7.3$ ps) (d) reflection from the right (red) end ($d_0 =4.0a$, $\sigma _x
=5$, $k_0 =0$, and $T_R =7.3$ ps).} \label{fig:gpc:CONTOUR}
\end{figure*}

\begin{figure*}[h t b p]\caption{(Color online) Comparison of the evolution of wavepackets of a undamped BO ($\Gamma
=0$, solid lines) in the ideal nanoparticle chian and of a damped BO ($\Gamma =0.08\omega _p $, dashed lines) in the silver nanoparticle chains
($d_0 =5.5a$,$\sigma _x =5$, $k_0 =0.8\pi /d_0 )$.} \label{fig:gpc:EVOL}
\end{figure*}

\newpage

\centerline{\includegraphics[width=0.8 \textwidth]{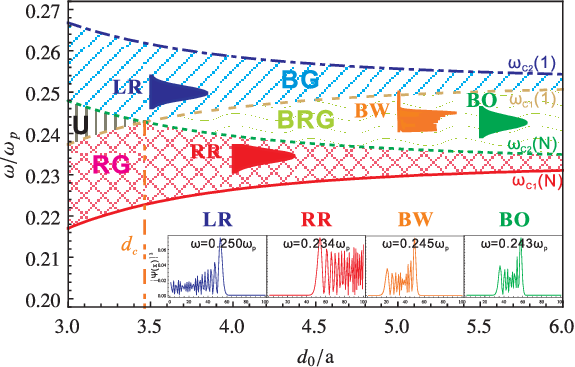}} \centerline{Fig.1./Zheng, Xiao, and Yu}

\newpage

\centerline{\includegraphics[width=0.6 \textwidth]{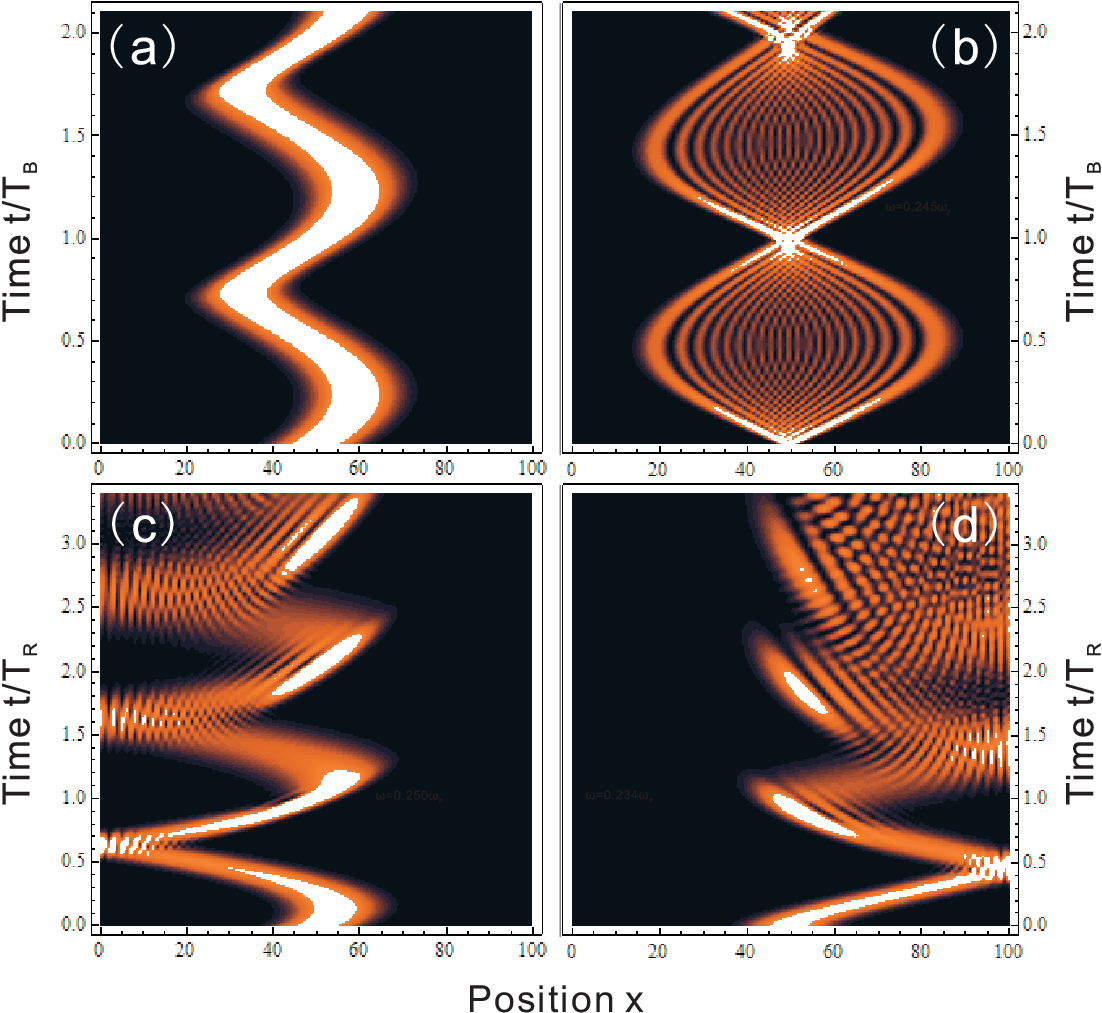}} \centerline{Fig.2./Zheng, Xiao, and Yu}

\newpage

\centerline{\includegraphics[width=0.6 \textwidth]{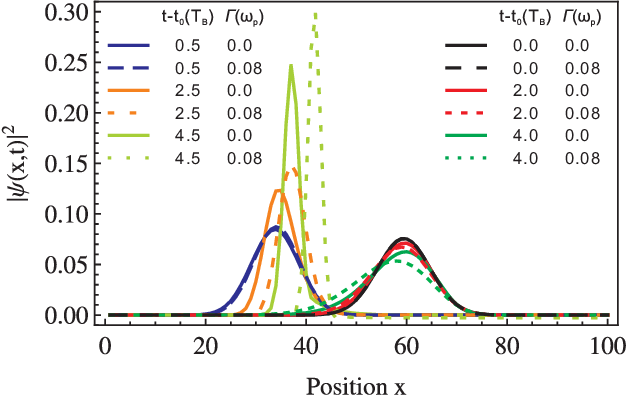}} \centerline{Fig.3./Zheng, Xiao, and Yu}

%\end{CJK*}
\end{document}